# The dynamics of individual nucleosomes controls the chromatin condensation pathway: direct AFM visualization of variant chromatin.

**Fabien Montel[1,2]\*, Hervé Menoni[3]\*, Martin Castelnovo[1,2], Jan Bednar[4,5], Stefan Dimitrov[6], Dimitar Angelov[3]¶ and Cendrine Faivre-Moskalenko[1,2]¶**

[1]Université de Lyon; Laboratoire Joliot-Curie, CNRS USR 3010, Ecole Normale Supérieure de Lyon, 46 allée d'Italie, 69364 Lyon Cedex 07, France ;

[2]Université de Lyon; Laboratoire de Physique, CNRS UMR 5672, Ecole Normale Supérieure de Lyon, 46 allée d'Italie, 69364 Lyon Cedex 07, France ;

[3]Université de Lyon; Laboratoire de Biologie Moléculaire de la Cellule, CNRS UMR 5239, Ecole Normale Supérieure de Lyon, 46 allée d'Italie, 69364 Lyon Cedex 07, France ;

[4]CNRS/UJF, Laboratoire de Spectrométrie Physique, UMR 5588, BP87, 140 Av. de la Physique, 38402 St. Martin d'Hères Cedex, France

[5]Charles University in Prague, Institute of Cellular Biology and Pathology, First Faculty of Medicine, and Department of Cell Biology, Institute of Physiology,Academy of Sciences of the Czech Republic, Albertov 4, 128 01 Prague 2, Czech Republic

[6]Université Joseph Fourier-Grenoble 1; INSERM
Institut Albert Bonniot U823, Site Santé, BP 170, 38042 Grenoble Cedex 9, France.

\* these authors contributed equally to this work.

¶ **corresponding authors :** C. Faivre-Moskalenko, Laboratoire Joliot-Curie, Ecole Normale Supérieure de Lyon, 46 Allée d'Italie, 69364 Lyon cedex 7, France. E-mail: cendrine.moskalenko@ens-lyon.fr, Tel : + 33 4 72 72 88 25, Fax : +33 72 72 80 80 ; or D. Angelov, Laboratoire de Biologie Moléculaire de la Cellule, Ecole Normale Supérieure de Lyon, 46 Allée d'Italie, 69364 Lyon cedex 7, France. E-mail: dimitar.anguelov@ens-lyon.fr, Tel : + 33 4 72 72 88 98, Fax : +33 72 72 80 80.

# Abstract

Chromatin organization and dynamics is studied in this work at scales ranging from single nucleosome to nucleosomal array by using a unique combination of biochemical assays, single molecule imaging technique and numerical modeling. We demonstrate that a subtle modification in the nucleosome structure induced by the histone variant H2A.Bbd drastically modifies the higher order organization of the nucleosomal arrays. Importantly, as directly visualized by AFM, conventional H2A nucleosomal arrays exhibit specific local organization, in contrast to H2A.Bbd arrays, which show "beads on a string" structure. The combination of systematic image analysis and theoretical modeling allows a quantitative description relating the observed gross structural changes of the arrays to their local organization. Our results strongly suggest that higher-order organization of H1-free nucleosomal arrays is mainly determined by the fluctuation properties of individual nucleosomes. **Moreover, numerical simulations suggest the existence of attractive interactions between nucleosomes to provide the degree of compaction observed for conventional chromatin fibers.**

# Introduction

The major elementary building blocks of chromatin (1) are known since a few decades to be the nucleic acid (DNA) bearing the genetic information, and the four different histones (H2A, H2B, H3, H4) arranged by pairs into an octamer. The first level of conserved organization of these molecules is the nucleosome, in which about 1,75 turns of DNA (147 bp) are wrapped around the histone octamer (2). The spatial organization of nucleosomal array in the presence of the linker histone H1 gives rise to several higher order structures of chromatin fiber, the first one being the so-called 30 nm chromatin fiber. Several different models for the 30 nm chromatin fiber structure were proposed in the literature (3-6). Chromatin is highly dynamic and numerous factors including nucleosome remodeling complexes, histone chaperones and histone variants are essential for maintaining its dynamics (7).

Histone variants are non-allelic isoforms of the conventional histones (1) and are expressed in a relatively low amount compared to their conventional counterparts suggesting that in addition to their structural role, they might have some specialized function (for a recent review see (8)). While all histones, except H4, possess their variants, H2A has the largest family of them (8). The histone variant H2A.Bbd ('*Barr body deficient*') belongs to the H2A family. It shares only 48% homology with its parental histone (9). H2A.Bbd is excluded from the X inactive chromosome of female vertebrate (10) and its localization in chromosome regions where the chromatin is acetylated suggests that H2A.Bbd could have a positive role in transcription (10).

A characteristic feature of the histone variant H2A.Bbd structure is that the residues that contribute to the nucleosome core particle (NCP) acidic patch are missing (9). In addition, it lacks the C-terminus characteristic of the H2A family as well as the end of the docking domain of H2A that was shown to be involved in the interaction of the H2A/H2B dimer with the $(H3/H4)_2$ tetramer (10, 11). Several types of experiments based on biochemical approaches or microscopy techniques have shown that less DNA is organized in H2A.Bbd nucleosomes compared to conventional nucleosomes (11). Moreover this sub-complexed nucleosomal structure is more dynamic (12, 13) and exhibits a weaker thermodynamic stability than the canonical nucleosome (12, 14). The more open structure of H2A.Bbd nucleosome was shown to facilitate the access of transcription factors (15) and base excision repair factors (16), which suggests that this variant nucleosome represents a lower physical barrier for chromatin associated processes.

By using a fusion protein Bbd.ddH2A (a H2A.Bbd chimera, in which the docking domain is replaced with that of conventional H2A), we were recently able to show that the docking domain is in part responsible for the specific properties of the H2A.Bbd mono-nucleosome (12). In addition, recent analytical centrifugation experiments demonstrated that H2A.Bbd nucleosomal arrays exhibited less compact structure in the presence of magnesium compared to that of conventional H2A arrays (17). This specific property of the H2A.Bbd arrays appeared to be determined by the lack of acidic patch in the H2A.Bbd histone octamer (17).

In this work, we use a combination of biochemical techniques, Atomic Force Microscopy and numerical modeling in order to visualize and compare quantitatively the structural and dynamic properties of reconstituted nucleosomal arrays with either conventional H2A or H2A.Bbd histone variant or chimeric Bbd.ddH2A protein. When combined with image analysis, AFM allows the detection of the position of each nucleosome within the analyzed chromatin coil. Subsequently, the 2D structure factor for each type of studied chromatin sample can be calculated, which enables us to probe the structure and dynamics of the chromatin at various scales ranging from the monomer size (~ 10 nm) to the whole fiber size (~ 100 nm). By comparing the experimental structure factors to those obtained for simulated chromatin arrays, we quantitatively relate the equilibrium parameters measured on the mono-nucleosome to the structural parameters describing the corresponding nucleosomal arrays and thereby unravel the individual nucleosome mechanical requirements for nucleosome array to fold into a compact fiber.

## **Results and Discussion**

Nucleosomal arrays reconstituted on 601 DNA repeats with different repeat numbers were used in the experiments. The 601 DNA sequence exhibits a high positioning signal that enables us to obtain both conventional and variant nucleosomes accurately positioned along the DNA at specific positions (18). In addition, our experimental conditions were restricted to low salt environment to prevent both variant octamer destabilization (19) and fiber-fiber interactions in high divalent salt concentration as used in the centrifugation studies (17).

Conventional H2A, histone variant H2A.Bbd and chimeric Bbd.ddH2A nucleosomal arrays were reconstituted by salt dialysis onto DNA fragments of 1, 1.8, 3 and 6.3 kbp containing 5, 9, 15 and 32 repeats of 601 positioning sequences respectively (repeat length 197 bp). Small amount of mono-nucleosomal sized 5S DNA was used as a competitor DNA in the chromatin reconstitution to achieve complete saturation of 601 chromatin. The quantity of competitor DNAs was low enough not to affect the AFM image and allowed tuning the histone DNA ratio in a fine manner to avoid aggregation.

**Biochemical characterization of the reconstituted arrays**

The reconstituted nucleosomal arrays were first characterized by microccocal nuclease digestion (Figure 1). The digestion pattern of the three types of 32 repeat nucleosomal arrays was very regular (more than 20 bands were visible at the lowest time of digestion). This strongly indicates saturation of the arrays under our experimental conditions and a precise positioning of the individual nucleosomes in each 601 repeat. It is to note that the H2A.Bbd and Bbd.ddH2A chromatin arrays are more accessible to the microccocal nuclease (Figure 1) suggesting that these arrays are less compact than the conventional ones. In addition, the digestion profile of the H2A.Bbd and Bbd.ddH2A chromatin fiber digestion patterns exhibit satellite bands around the main band, in agreement with the digestion pattern of similar mono-nucleosomes (see supplemental figure S2). This is consistent with our previous observation of larger opening fluctuations of these mono-nucleosomes compared to the conventional ones.

**AFM visualization of conventional and variant nucleosomal arrays**

Atomic force microscopy has been frequently applied to study conventional nucleosome array ((20) and references within) and our goal is to extend this type of approach to variant chromatin fibers where direct measurements are rather scarce (21). The chromatin fibers are imaged in tapping mode AFM in air. The chromatin samples were deposited on APTES functionalized mica surfaces; this type of self-assembled monolayer has been shown to trap bio-molecules on the surface into a configuration in 2D that is the projection of the 3D equilibrium configuration of the molecule in solution (22). This adsorption protocol has two main advantages: it preserves the structure of the fiber as it was in solution (a low salt buffer in our case, 10 mM Tris, 5 mM NaCl and no $MgCl_2$), and it does not require the use of a biochemical glue such as glutaraldehyde, that may lead to artifacts in the visualization process through the cross-linking of two amine groups (mainly on the lysines of histone tails) (23).

Typical AFM images obtained for 32 repeat chromatin fibers are presented in figure 2. The good saturation of nucleosomal arrays can be observed for the 3 types of chromatin: conventional H2A (Fig.2a to Fig.2d), variant H2A.Bbd (Fig.2e to Fig.2h) and chimeric Bbd.ddH2A protein (Fig.2i to Fig.2l), in agreement with Mnase digestion. It should be stressed that the whole set of AFM images shows only 2D chromatin arrays (see the height of the image). Several chromatin fibers too close to be separated were sometimes observed (Fig.2a for example), but they were rejected after the automated image analysis.

In the case of the conventional nucleosomes (Fig.2a to 2d), some clear compaction is observed on the AFM images. The compaction of the fiber is such that the DNA trajectory in between nucleosomes (linker DNA ~ 50 bp) is not always visible. In most of the AFM images obtained for the 32 conventional nucleosome fibers, one can observe a typical 2D zig-zag structure in distinct parts of the fibers. This organization and compaction level of the conventional nucleosome fibers was also observed in electron cryo-microscopy (EC-M) imaging (supplemental figure S3). The strong positioning signal of the 601 DNA sequence is likely to favor such a regularity in the nucleosome organization (24). Therefore, by using two independent microscopy approaches we have obtained essentially the same results, suggesting that our AFM imaging experiments are artifact-free.

Conversely, the arrays of H2A.Bbd variant nucleosomes exhibit a more relaxed 'beads-on-a-string' type structure. Finally, the chimeric histone variant Bbd.ddH2A also affects the nucleosomal array compaction, and leads to an open relaxed structure, similar to the H2A.Bbd variant fiber, thus confirming the interpretation of Mnase digestion pattern.

**Conventional and histone variant nucleosomal arrays exhibit different local and global properties**

Quantitative information can be extracted from the AFM image analysis using a homemade Matlab script (see *Materials and Methods*). Unlike the mono-nucleosome image analysis procedure used in our previous study (13), a simple height criterion is not sufficient to faithfully detect each nucleosome position within the compact conventional fiber. Therefore, we have implemented an algorithm that identifies local curvature maxima, thereby enabling to detect the position of the NCP centroid with a sub-nanometer precision. For each fiber 'object' identified, the script measures several parameters of interests (see *Materials and Methods* section). For the calculation of these various parameters, only the fibers with a number of nucleosomes in agreement with the expected value for each DNA construction were selected ($4 \leq N_{nuc} \leq 5$ for 5 repeats of 601 positioning sequences, $7 \leq N_{nuc} \leq 10$ for 9 repeats, $12 \leq N_{nuc} \leq 17$ for 15 repeats and $25 \leq N_{nuc} \leq 35$ for 32 repeats) and this at the expense of our statistical sampling. These criteria allow elimination of over- and sub-saturated fibers as well as the 'connected' fibers from the data analysis.

We discuss here only the most relevant parameters: $R_g$ the radius of gyration, $R_H$ the hydrodynamic radius, $N_{nucs}$ the number of nucleosomes in the selected fiber, $d_{1st\_neighbour}$, the distance to its nearest neighbour for each nucleosome and $d_{inter\_nuc}$, the average inter-nucleosomal distance within the fiber. The mean value of each quantity has been estimated for each type of reconstituted fiber (conventional or variant, 5, 9, 15 or 32 repeats) and the complete data are summarized in Table 1. The results for two relevant representative parameters are represented in figure 3: the nearest neighbour distance (Fig.3a) which is a local parameter characterizing the fiber organization, and the radius of gyration (Fig.3b) which is a global one.

The fiber configuration can be characterized at the monomer scale by calculating for each nucleosome the distance to its nearest neighbour. The nearest neighbour distance distribution obtained for each type of 32 repeat nucleosomal fibers is plotted in figure 3a. The conventional nucleosome nearest neighbour distance is centered on $<d_{1st\_neighbour}>= 20.1 \pm 0.3$ nm and the value found for each DNA template (5, 9, 15 or 32 repeats) is very close (see Table 1) showing that the local organization of the conventional fiber is similar for several saturated DNA template lengths. For variant nucleosomal fibers (H2A.Bbd and Bbd.ddH2A), the nearest neighbour distance distribution is markedly broaden and asymmetric. This reflects a larger tendency of nearest neighbour nucleosomes in the case of variants to be less localized, and therefore a smaller degree of local compaction of the fiber.

The data measured at the local scale on our reconstituted chromatin can also be compared to previous AFM measurements on native chromatin. In particular, the nearest neighbour distance and the average inter-nucleosomal distance found for conventional chromatin are consistent with data from Kepert *et al*. (25). In this study, a mean value of $17.6 \pm 0.1$ nm for the nearest neighbour distance and $27.6 \pm 0.6$ nm for the inter-nucleosomal distance are found for native chromatin fibers extracted from HeLa cells, depleted from linker histone H1. Despite the difference in the origin of chromatin studied and the deposition conditions for AFM imaging, the similarities of these results show that the features of extracted data are intrinsically relevant of chromatin structure.

At a global scale, this difference in compaction is also observed through the comparison of typical radii (radius of gyration, hydrodynamic radius) between conventional and variant fibers (Fig 3b and Table 1). The mean radius of gyration of conventional fibers with 32 nucleosomes (on average) is $R_{g\_H2A} = 71.8$ nm, while the same mean radius for the variant fiber is $R_{g\_H2A.Bbd} = 88.1$ nm. The compaction of conventional fibers with respect to H2A.Bbd variant fibers has already been measured at this global level by Zhou *et al*. (17) for chromatin with 12 nucleosomes. Our results for similar fibers (with 9 nucleosomes per fibers, see Table 1) are qualitatively consistent with theirs (17), the relative deviation being easily explained by different buffer conditions and the difference between 2D and 3D fibers. Nevertheless, the use of image analysis to compute global parameters like radii of gyration allows us to go beyond the average value of radii and to obtain its full distribution on the given set of analyzed fibers. Again, the larger width of this distribution in the case of variant fibers (cf Fig.3b) is consistent with a smaller degree of fiber compaction. However, further investigation of radius of gyration scaling with the number of nucleosomes is hampered due to the limited range of scales experimentally accessible.

In summary, we have shown that both at the local and global scales the variant chromatin fiber is statistically more open and less organized than the conventional one.

**2D structure factors allows quantifying the compaction of conventional fibers with respect to the variant fibers**

In order to gap our observations between the local and the global scale of the fiber, we computed two-dimensional structure factors out of inter-nucleosome distances measurements, following the procedure described in the *Materials and Methods* section. The use of structure factors has two main advantages: first the quantification of the fiber structure at different length scales (26), and second the comparison between experimental and simulation results.

Using the distances between each nucleosome for each fibers extracted from the image analysis, we compute a 2D structure factor ($S(q)$). This quantity bears useful information on the structure of the observed objects at different scales, ranging from the nucleosome scale to the global fiber scale. The 3D structure factors are usually obtained by various Small Angle Scattering techniques (neutrons, X-rays, or light). Within our experimental setup, **computing artificially a structure factor from real images may not make sense at first glance**, but it turns out to be an extremely powerful tool to quantitatively compare experimental results and numerical simulations at various length scales, as it is discussed below.

The experimental 2D structure factors are conveniently represented as Kratky plots ($q^2 S(q)$ vs $q$) (27). Within such a representation, a simple semi-flexible polymer (for example DNA) will exhibit 3 regimes : at low $q$ (i.e. for distances larger than the radius of gyration $R_g$ of the coil), $q^2 S(q)$ increases as a function of $q$ (Guinier regime, where $S(q)$ decays exponentially), then for $R_g < q <$ monomer size, there is a plateau corresponding to a Gaussian chain regime (where $S(q)$ scales as $q^{-2}$), and finally for large $q$ (i.e. sizes smaller than the monomer size) $q^2 S(q)$ increases linearly with $q$ (rigid rod regime $S(q) \propto q^{-1}$). An additional peak may eventually appear in the Kratky plot representation, that is associated to a structure that is more compact (or organized) at an intermediate scale between monomer and coil size, than a Gaussian chain. This peak is a typical signature of intramolecular partial compaction, as it has been shown recently to monitor folding/unfolding transition in RNA and proteins (27, 28).

Experimentally, conventional fibers with 9 and 32 nucleosomes exhibit these 3 regimes with a significant peak associated to some degree of compaction in the structure, while variant fibers with the same nucleosome numbers do not (Fig. 4a). As we already mentioned, this maximum in our experimental data is the signature of the tendency to form locally some ordered (zig-zag) configuration of conventional nucleosomes, as can be observed directly on many images of fibers (Fig. 2a to 2d) or exhibited by simulation results on highly ordered fibers (supplemental data, figure S5). On the contrary, the absence of any significant peak in the Kratky plot of variant fibers indicates an organization of the whole chain that is closer to a random walk or Gaussian chain. Interestingly, the structure factor of chimeric Bbd.ddH2A fiber is closer to the one of Bbd.H2A fiber, in agreement with image snapshots shown on figure 2.

In order to gain more insights into the interpretation of these structure factors, we developed simple simulations of 2D chromatin fibers, as it is described in the *Materials and Methods* and in the *Supplemental Material*. Using the experimental distribution of DNA complexed length for both conventional and variant mono-nucleosomes as an input, we were able to generate different set of representative conformations, from which we calculated 2D structure factors. For each type of simulated chromatin fibers, we averaged over 500 chains in order to ensure for statistical reliability of the Kratky plots. Focusing first on the conventional and variant fiber data (Fig. 4b and 4d), a remarkable agreement can be observed between the experiments and the simulations once an appropriate excluded volume is chosen for all nucleosomes. In particular, the low-$q$ regime, *i.e.* at the fiber scale, is well described within our model. This means that using a single model for chromatin fibers, together with two different distributions of nucleosome complexation length representing different histone compositions, it is possible to capture quantitatively the relevant features of the observed fiber conformations. The only adjustable parameter for these simulations is the choice of excluded volume distance ($d_{ev}$) between nearest nucleosomes, whose optimal value is found to be $d_{ev} \sim$ 17 nm. This value is consistent with both the experimental average nearest neighbour distance, and the typical excluded volume due to the presence of histone tails (29).

Remarkably, the experimental radius of gyration matches the peak (or inflection point) position in the Kratky plots as evidenced in figure 4c. A closer inspection of the structure factors for conventional fibers with different number of nucleosomes (5, 9, 15 and 32) at moderate-q regime ($10^{-2} < q < 10^{-1}$) reveals however some quantitative discrepancies (Fig. 4c). These discrepancies between the experiment and the theory become more evident with increasing number of nucleosomes (15 or 32 nucleosomes) in the array. Indeed, further analysis of simulations with pure excluded volume interactions (Fig. 4b) shows that although the relative "rigidity" of conventional nucleosomes seems to be enough to produce some compaction or structuring of the array for 5 or 9 nucleosomes, it is not able to compact larger number of nucleosome (15 or 32 nucleosomes). This means that some physical ingredient like nucleosome attractions favoring compaction over a larger range of scales is missing in order to reproduce the experimental structure factors.

In order to qualitatively test this assumption, we extended our simulations to include effective attractions between nucleosomes. This was simply achieved as a first approximation by increasing the acceptance rate in the process of chain construction for nucleosome distances close to solid contact relatively to larger distances. This generates chains that exhibit a stronger degree of compaction. If a large number of chains is generated this way (500 chains), the structure factor shows now a significant peak in the Kratky representation compared to the same simulation with pure excluded volume interactions, in qualitative agreement with the experimental results. As the experimental Kratky plots were obtained from a rather limited set of chromatin chains, we observe interestingly in the simulation, that lowering the statistics of chain generation to values similar to the experimental results (roughly 50 chains) produces structure factors remarkably close to the experimental one (see figure 4c).

The quantitative agreement between our simulations and our AFM data shows that the only input of the mono-nucleosome DNA complexation length distribution, or equivalently the mean opening angle and the nucleosome flexibility, is sufficient to describe the multi-scale behaviour of conventional and variant chromatin arrays. To discriminate the role of each ingredient (angle or flexibility), the results of the chimeric variant Bbd.ddH2A chromatin can be used. Indeed, as it was mentioned in the introduction, the complexation length distribution of DNA on chimeric Bbd.ddH2A mono-nucleosomes has roughly the same average value (opening angle) as the conventional one, and the same large width (flexibility) as the variant one. Since structure factors of the chimeric fiber with either 9 or 32 nucleosomes are closer to the one of the variant, one can argue that the fluctuations of DNA wrapped length has a larger influence in determining higher-order chromatin structure than the average wrapped length. This is further confirmed by our simulations as shown in supplemental Fig. S7, and leads to the important following conclusion: the nucleosome flexibility seems to be the main ingredient to the route of chromatin fiber compaction. The picture arising from this study is that chromatin whose nucleosomes are more flexible than conventional one is unable to form spontaneously a higher order structure. Indeed, too large fluctuations might impede the nucleosomes to feel the attractive interactions with its neighbours or to display the favorable configuration for fiber formation (30).

Let us now discuss our findings in regards to the results of Zhou *et al.* (17). They have seen that recovering the acidic patch of H2A on the H2A.Bbd histone is necessary for compacting the H2A.Bbd fiber, but not enough to recover the full level of compaction of the conventional chromatin without $MgCl_2$. The authors hypothesize that interactions between the acidic patch on the surface of H2A and the H4 tail of the same nucleosome are responsible at the microscopic level for the ability of chromatin to fold into a compact fiber. Within the framework of our mechanical view, the origin of the difference in chromatin compaction arises from the flexibility of the nucleosome at the individual scale. The loss of interaction of H4 histone tails with the acidic patch on the nucleosome surface is a good candidate to explain the increased flexibility of the variant H2A.Bbd nucleosome observed at the mono-nucleosome level.

Therefore, our mechanical model of chromatin organization allows linking the microscopic origin of the H2A.Bbd variant increased flexibility to the formation of the higher order chromatin structure. In this context, post-translational modifications of histone tails could also induce a change in nucleosome flexibility that might explain the observed regulation of chromatin compaction (31).

## **Conclusions**

In this work, we investigated quantitatively the relation between mono-nucleosomes intrinsic properties for different histone contents with the higher-order structure of chromatin fibers. This was achieved by the unique combination of biochemical methods, AFM visualization and numerical simulations. The comparison of fiber's structures for conventional, H2A.Bbd variant and Bbd.ddH2A chimeric nucleosome content probed by all three methods lead to the following conclusions: there is a direct connection between DNA complexation length distribution on mono-nucleosomes and the structure of nucleosomal array. More precisely, the width of this distribution, or equivalently the spontaneous tendency of nucleosome to unwrap more or less easily, turns out to be a major determinant of higher-order structure as observed through AFM. Moreover, the use of simulations allowed highlighting the role of attractive interactions among nucleosomes in providing the observed degree of compaction for conventional fibers.

These results have some important biological implications. They strengthen the idea that the ability of H2A.Bbd histone variant to modify the structural and dynamic properties of the mono-nucleosome provides a regulation pathway for DNA accessibility within the chromatin fiber.

In a more general context, our results suggest that any process likely to modify mono-nucleosome dynamics (like a transcription factor binding, chromatin remodeling or post-traductional histone modifications) can potentially induce a modification of a higher order chromatin structure. They highlight the deep role of fluctuations at the nucleosome scale for the whole chromatin organization. Therefore, a next step would be to study how localized flexibility defect generated by presence of a single variant nucleosome, would propagate to neighbouring nucleosome creating a locally opened chromatin structure.

## Materials and Methods

**Preparation of DNA fragments**

The DNA fragments containing 5, 9, 15 or 32 repeats of 601 sequence (197 bp) were constructed essentially as described in (32). The long DNA fragments for chromatin reconstitution were gel or PEG purified as described in (32).

**Protein purification, nucleosome and chromatin reconstitution**

Recombinant Xenopus laevis full-length histone proteins were produced in bacteria and purified as described (33). For the H2A.Bbd protein and the H2A.Bbd-ddH2A mutant (H2A domain from M1 to I80 fused to H2A.Bbd domain from T84 to D115), the coding sequences were amplified by PCR and introduced in the pET3a vector. Recombinant proteins were purified as previously described (15).

Chromatin reconstitution was performed by the salt dialysis procedure (34). A low quantity (<~ 10%) of competitor 5S DNA fragments was added to avoid over-saturation of the nucleosomal array.

**Biochemical analysis**

Micrococcal nuclease digestion was performed at 8 U/ml at room temperature for indicated times in 10 mM Tris, pH 7.4, 1 mM DTT, 25 mM NaCl, 5% glycerol, 100 µg/ml BSA, 1.5 mM $CaCl_2$ and 100 µg/ml of plasmid carrier DNA. The digestion was stopped by adding 20 mM EDTA, 0.1% SDS, 200 µg/ml Proteinase K (30 min at 45°C). DNA was then extracted and run on a 10% native acrylamide bisacrylamide (1/29 w/w) gel or 1.4% agarose gel.

**Atomic Force Microscopy and surface preparation**

For the AFM imaging the conventional and variant nucleosomal arrays were immobilized onto APTES-mica surfaces. The functionalization of freshly cleaved mica disks (muscovite mica, grade V-1, SPI) was obtained by self-assembly of a monolayer of APTES under Argon atmosphere for 2 hours (35). A 5 µl droplet of the chromatin solution in low salt buffer (10mM Tris pH = 7.4, 1 mM EDTA and 5 mM NaCl) was deposited onto the APTES-mica surface for 1 min, rinsed with 1 mL of milliQ-Ultrapure© water and gently dried by nitrogen flow. The samples were visualized by using a Nanoscope III AFM (Digital Instruments™, Veeco, Santa Barbara, CA). The images were obtained in Tapping Mode in air, using Diamond Like Carbon Spikes tips (resonant frequency ~150 kHz) at scanning rates of 2 Hz over scan areas of 1 µm wide.

**Image analysis**

The parameters of interest were extracted from the AFM images using a homemade MATLAB © (The Mathworks, Natick, MA) script essentially based on morphological tools such as binary dilatation and erosion (36), and height/areas selections. The aim of the first two steps of this algorithm was to select relevant objects:

1. In order to remove the piezoelectric scanner thermal drift, flattening of the image is performed. The use of a height criteria (h>0.5nm where h is the height of the object) allows us to avoid the shadow artifact induced by high objects on the image.

2. Building of a binary image using a simple thresholding (h > 0.25 nm where h is the height of the object) followed by selection of the binary objects in the good area range (X < A < Y nm² where A is the area of the object, X and Y depends on the number of repeats).

These two steps lead to the selection of binary objects whose area is between for example for X = 5000 nm² and Y = 15000 nm² for a five repeats of 601 positioning sequence and corresponds in the AFM image to a group of connected pixels which minimum height is more than 0.25 nm.

The next steps correspond to the characterization of these objects done automatically for each selected chromatin fiber

3. Measurement of the fiber projected total area, $A_{tot}$, (number of pixels above the noise threshold (0.25nm) for an object in the good area range)

4. Segmentation of the NCPs by selecting regions exhibiting a local curvature below -0.01 nm$^{-1}$ and a size larger than 20 nm$^2$.

5. Detection of the NCP centroid by extracting the center of mass for each NCP and determination of the number $N_{nucs}$ of NCPs in this fiber.

6. Measurement of Euclidian distances ($d_{ij}$) between centroids of NCPs $i$ and $j$, for $i$ and $j = 1$ to $N_{nucs}$ using distances.

7. Extraction of the first two principal components of the 2D series defined by NCP centroids. Determination of the major and minor axis of the ellipse defined by the two principal directions and the value of the associated semi-major axis $a$, and semi-minor axis $b$.

8. Estimation of $C_{2D}$, the fiber 2D compacity by calculating the ratio between the fiber projected area $A_{tot}$ (estimated in step 3) and the ellipse area $A_{Ellipse} = \pi*a*b$.

9. Determination of the distance to its nearest neighbor ($d_{1st\_neighbour}$) for each NCP.

10. Estimation of the characteristic distance between nucleosomes by computing

$$d_{inter\_nucs} = 2\sqrt{\frac{A_{total}}{\pi N_{nucs}}}.$$

11. Calculation of the radius of gyration, $R_g$, defined as the mean square distance to the center of mass for all NCP centroids that belong to one object, $R_g^2 = \frac{1}{N^2}\sum_{i=1}^{N}\sum_{j=1}^{N}d_{ij}^2$.

12. Calculation of the hydrodynamic radius, $R_h$, defined as $R_H^{-1} = \frac{1}{N(N-1)}\sum_{i=1}^{N}\sum_{j=1,j\neq i}^{N}\frac{1}{d_{ij}}$, where $d_{ij}$ is the distance between centroids of NCPs $i$ and $j$ (calculated in step 6) and N the total number of nucleosome in the fiber.

The steps 4 and 5 lead to quick and robust measurements. Indeed the combined use of local curvature, area threshold and center of mass to locate NCP centroid lead to a sub-nanometer resolution in the X/Y positions and exclude compactly bent DNA from being considered as a candidate NCP.

For each estimated global or local structural parameter, the error on the mean value of the distribution is estimated as $\sigma/\sqrt{N}$, where $\sigma$ is the standard deviation of the distribution and $N$ the total number of objects.

**Structure factor calculations**

From the image analysis previously described, it is possible to extract distances $d_{ij}$ between each nucleosomes center on each analyzed chromatin type. Using these data, the 2D structure factors (istropically averaged) are calculated as follows:

$$S(q) = \frac{1}{N} \sum_{i=1}^{N} \sum_{j=1}^{N} J_0(q \cdot d_{ij})$$

where $J_0$ is the zeroth order Bessel function of the first kind. The analysis of structure factors can benefits from many different representations developed over the last 50 years within the field of polymer physics. In particular, the Kratky plot representation ($q^2 S(q)$ vs $q$) of a structure factor is a convenient way to highlight a locally compact structure, as it is shown by its recent use in the characterization of protein or RNA folding/unfolding by Small Angle X-Ray Scattering (SAXS) (27). Indeed within such a representation, any peak in the curve is associated to a structure that is more compact than the equivalent random walk or Gaussian chain. In the case of a Gaussian chain, the structure factor scales like $q^{-2}$, while such for a compact state the structure would scale like $q^{-\alpha}$ where $2 < \alpha \leq 4$.

**Numerical simulations**:

The purpose of numerical simulations performed in this work is to extend the analysis of experimental data obtained by AFM visualization of chromatin fibers. We describe more precisely in this section the rules of the simulations. Our 2D model of H1-depleted chromatin fibers has essentially four major ingredients:

(i) a basic mechanical model taking into account the geometrical relation between DNA complexed length within each nucleosome and linker length between consecutive nucleosomes

(ii) the possibility to use as an input the experimental distribution of mono-nucleosome opening angles obtained in our previous work (13), through the equivalent DNA complexation length distribution

(iii) the excluded volume between nucleosome core particles (NCP)

(iv) eventually some short range attractive interactions between NCP

The building blocks of the model are hard disks representing NCP and straight linkers. The first step is essential in providing realistic 2D positioning distributions of consecutive nucleosomes. The relevant exact geometrical relationships are summarized on supplemental figure S4. Each chain is then constructed as follows. We first decided to construct the chain of $N$ nucleosomes by placing the nucleosomes sequentially: this assumption is supposed to mimic the process of deposition of fibers on the surface starting from one of their end. Once the $i^{th}$ nucleosome is placed, the position of the next one is determined by choosing first a trial angle $\theta$ from the distribution of DNA complexation length of mono-nucleosomes observed experimentally. Any deviation from the canonical value of $\theta = \theta_0$ is translated into linker length variation according the relations in figure S4. It should be noted that this relation assumes that the linker variation are done in a forward way. Any piece of chain already constructed is immobilized for the rest of the construction process. Once the opening angle and linker length are known, the putative position of the $(i+1)^{th}$ nucleosome is known. If the position does not overlap with any previous NCP with effective diameter $d_{ev} = 17$ nm (the most optimal choice), the position is accepted and the computation proceeds towards the next step, while upon NCP overlap a new angle $\theta$ is repeatedly generated until successful position has been found.

The specificity of the model with respect to histone content is made by choosing as an input, different DNA complexation length distributions for different histone content (conventional and variant). We previously characterized these DNA complexation length distribution on conventional and variant mono-nucleosome by measuring its mean $<L_c>$ and width $\sigma_{Lc}$. In particular, we have shown that (12):

- for the conventional H2A nucleosome : $<L_{c\_H2A}> = 146 \pm 1$ bp, and $\sigma_{Lc\_H2A} \sim 20$ bp,

- for the variant H2A.Bbd nucleosome, the distribution is enlarged and shifted toward lower $L_c$ value : $<L_{c\_H2A.Bbd}> = 127 \pm 2$ bp, and $\sigma_{Lc\_H2A.Bbd} \sim 35$ bp

- for the chimeric Bbd.ddH2A nucleosome, the mean value is shifted back close to the conventional nucleosome wrapped length distribution but its width remains large : $<L_{c\_Bbd.ddH2A}>$ = 143 ±2 bp, and $\sigma_{Lc\_Bbd.ddH2A}$ ~ 35 bp.

Using these rules, a set of chains is then generated. The number of nucleosomes per fiber was chosen to be 5, 9, 15 and 30 for the different constructs, so that this number matched with the average number of nucleosome per fibers. Our simulation therefore neglects the effect of polydispersity in the number of nucleosome per fibers. From the chains generated this way, it is possible to compute all the characteristic quantities discussed in the paper: radius of gyration, hydrodynamic radius, nearest neighbour distribution, pairwise distance distribution, and structure factors. Representative snapshots of simulated fibers and the corresponding Kratky plots of the structure factors are shown in Figure S5.

## **Acknowledgements**

The authors acknowledge Dr Daniela Rhodes for the kind gift of the 32 repeat 601 array DNA template.

We thank Sajad Syed for his help with the 15 repeat chromatin reconstitution used in the cryo-microscopy experiments.

This work was supported by the Contrat Plan-Etat Région "Nouvelles Approches Physiques des Sciences du Vivant". D.A. acknowledges the Association pour la Recherche sur le Cancer for financial support. S.D acknowledges La Ligue Nationale contre le Cancer (Equipe labellisée La Ligue).

J.B. acknowledges the support of the Grant Agency of the Czech Republic (Grant #304/05/2168), the Ministry of Education, Youth and Sports (MSM0021620806 and LC535) and the Academy of Sciences of the Czech Republic (Grant #AV0Z50110509).

# Figure captions

**Figure 1**: Microccocal nuclease digestion kinetics of 32 mer chromatin. Identical amount of conventional H2A, variant H2A.Bbd and chimeric Bbd.ddH2A chromatin were digested with 8U/ml of microccocal nuclease for the indicated times. The reaction was stopped by addition of 20 mM EDTA and 0.1 mg/ml proteinase K, 0.1% SDS. DNA was isolated and run on a 1.4% agarose gel. 1 kbp M, marker DNA. The molecular mass of the fragments is indicated on the left part of the figure.

**Figure 2**: Typical set of AFM topographic images obtained in Tapping Mode in air for nucleosome arrays reconstituted on 32 repeats of 601 positioning DNA sequences (repeat length 197 bp) with (a to d) the conventional histone H2A, (e to h) the histone variant H2A.Bbd and (i to l) the chimeric variant histone Bbd.ddH2A.

**Figure 3** : Local and global parameters as measured with automated computer analysis of the AFM images. (a) Nearest neighbour distance distribution for conventional H2A (black line), variant H2A.BBd (dark gray line) and chimeric Bbd.ddH2A (light gray line) nucleosomal arrays reconstituted on the 32 repeats of 601 DNA fragment; (b) Radii of gyration for conventional H2A (black), variant H2A.BBd (dark gray) and chimeric Bbd.ddH2A (light gray) nucleosomal arrays reconstituted on the 9 and 32 repeats of 601. The radius of gyration distribution is conveniently displayed as a box plot, where the horizontal inner line corresponds to the median value. The lower and upper bounds of the box point respectively the first and last quartiles of the distribution. Notches represent a robust estimate of the uncertainty about the medians for box-to-box comparison.

**Figure 4** : Structure factors analysis of conventional and variant chromatin fibers. (a) Experimental Kratky plots for nucleosomal arrays reconstituted on 32 and 9 repeats respectively for conventional H2A (cyan and dark blue), variant H2A.Bbd (light green and dark green) and chimeric Bbd.ddH2A (orange and red). (b) Experimental Kratky plots (solid lines) for conventional H2A nucleosomal arrays reconstituted on 5 (cyan), 9 (light blue), 15 (blue) and 32 (dark blue) repeat 601 DNA fragments. Corresponding Kratky plots of structure factors averaged over 500 simulated nucleosomal arrays (dotted lines) with the parameters of the conventional H2A mono-nucleosome (average angle $\theta = 0.5\ \pi$ and flexibility $\sigma_\theta = 0.4\ \pi$)

for either 5 (cyan), 9 (light blue), 15 (blue) or 30 (dark blue) repeats. The $q$ value corresponding to the mean $R_g$ value estimated using image analysis is reported as a vertical dashed line for each array length (c) Kratky plots of structure factors for 30 repeat nucleosomal arrays simulated with the parameters of the conventional H2A mono-nucleosome ($\theta = 0.5\ \pi$ and $\sigma_\theta = 0.4\ \pi$) and only excluded volume (dark blue dotted line, averaged over 100 chains) or excluded volume and attraction (purple dashed line, averaged over 50 chains) in the model are compared with the experimental kratky plot for conventional H2A nucleosomal array of 32 repeats (dark blue solid line). (d) Experimental Kratky plots (solid lines) for variant H2A.Bbd nucleosomal arrays reconstituted on 9 (light green) and 32 (dark green) repeat 601 DNA fragments, and corresponding Kratky plots for structure factors averaged over 500 simulated nucleosomal arrays (dotted lines) with the parameters of the variant H2A.Bbd mono-nucleosome ($\theta = 0.7\ \pi$ and $\sigma_\theta = 0.7\ \pi$) for either 9 (light green) or 30 repeats (dark green).

**Table 1 caption** : Various parameters extracted from the automated images analysis describing the local and global conformation of the conventional H2A, variant H2A.Bbd et chimeric Bbd.ddH2A chromatin fibers of various sizes. Error is calculated as the standard error on the mean ($\frac{\sigma}{\sqrt{N}}$ where $\sigma$ is the standard deviation on the mean and $N$ the number of events in the distribution.

**Table 1** :

|  | number of repeats | total $N_{fibre} / N_{nucl}$ | mean $N_{nucl}$ / fibre | Radius of gyration (nm) | Hydrodynamic radius (nm) | nearest neighbour distance (nm) | inter-nucleosomal distance (nm) |
|---|---|---|---|---|---|---|---|
| **Conventional H2A** | 5 | 1335 / 6185 | 4,63 ± 0,01 | 22,1 ± 0,1 | 28,8 ± 0,1 | 21,2 ± 0,1 | 26,8 ± 0,1 |
|  | 9 | 261 / 2338 | 8,96 ± 0,06 | 32,6 ± 0,4 | 35,5 ± 0,3 | 19,8 ± 0,1 | 26,4 ± 0,1 |
|  | 15 | 551 / 8177 | 14,79 ± 0,06 | 53,7 ± 0,5 | 50,2 ± 0,3 | 21,5 ± 0,1 | 28,7 ± 0,1 |
|  | 32 | 54 / 1629 | 30,2 ± 0,4 | 71,8 ± 2,2 | 62,5 ± 1,1 | 20,1 ± 0,3 | 28,5 ± 0,3 |
| Variant H2A.Bbd | 9 | 132 / 1116 | 8,45 ± 0,08 | 46,0 ± 1,1 | 47,6 ± 0,8 | 26,2 ± 0,3 | 30,4 ± 0,3 |
|  | 32 | 19 / 593 | 31,2 ± 0,7 | 88,1 ±4,9 | 73,2 ± 2,9 | 21,7 ± 0,3 | 30,0 ± 0,7 |
| Chimeric Bbd.ddH2A | 9 | 112 / 995 | 8,9 ± 0,1 | 50,2 ± 1,1 | 50,9 ± 0,9 | 26,8 ± 0,3 | 32,8 ± 0,4 |
|  | 32 | 28 / 795 | 28,4 ± 0,5 | 95,3 ± 4,4 | 79,7 ± 2,6 | 24,2 ± 0,3 | 29,9 ± 0,5 |

**Montel *et al.* (2008)  Figure 1**

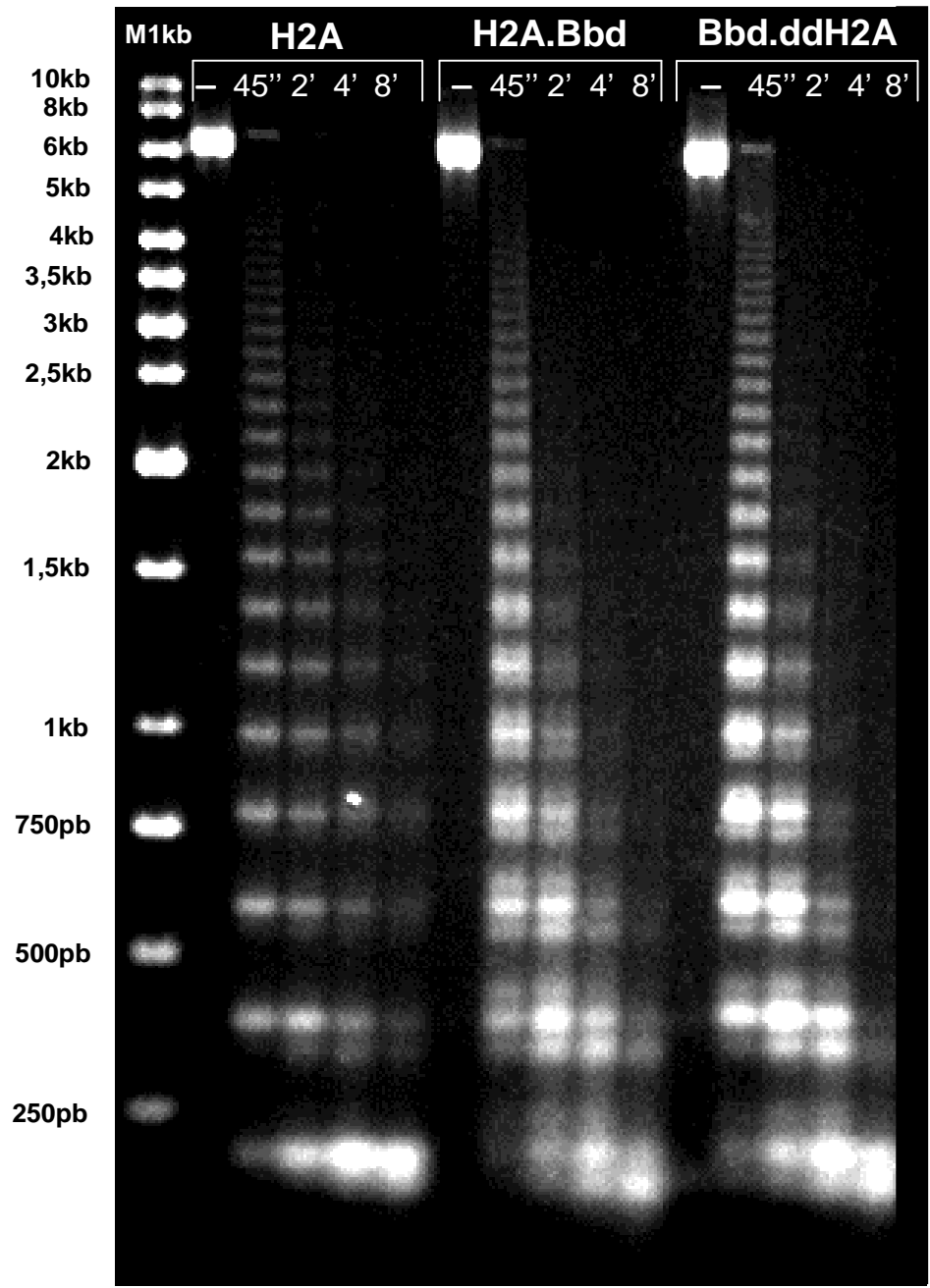

**Montel *et al.* (2008) Figure 2**

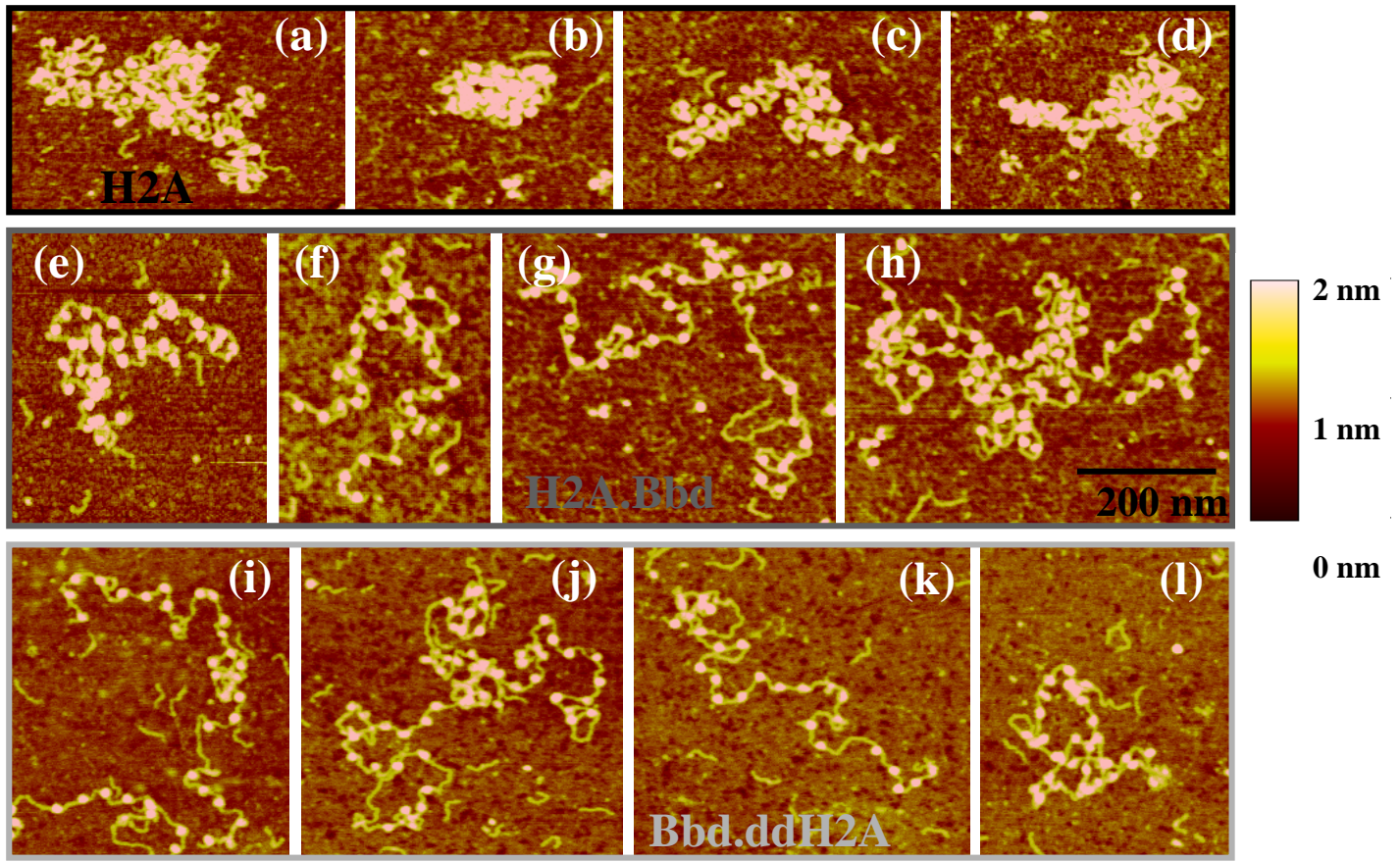

**Montel *et al.* (2008) Figure 3**

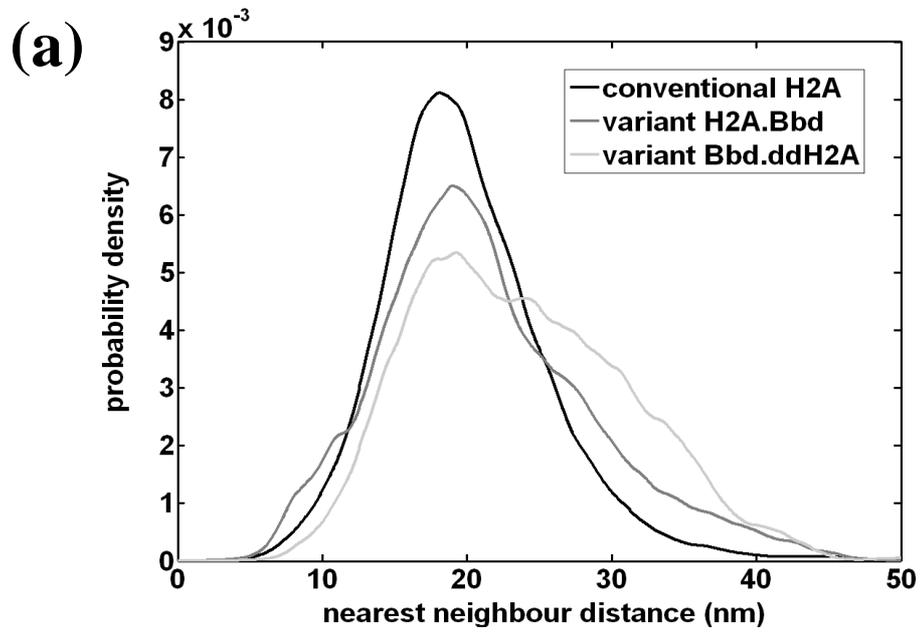

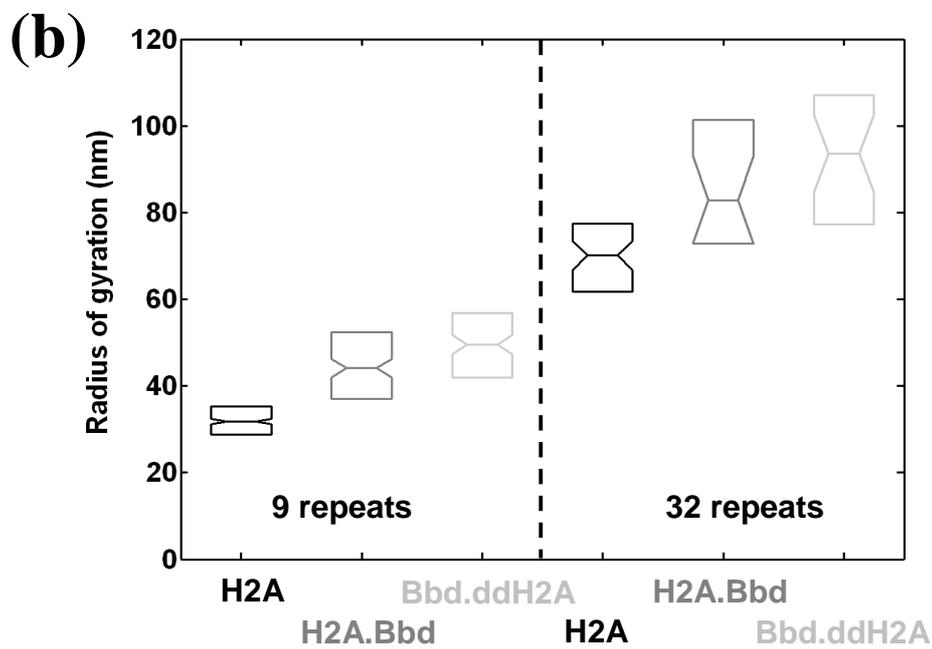

# Montel *et al.* (2008) <u>Figure 4</u>

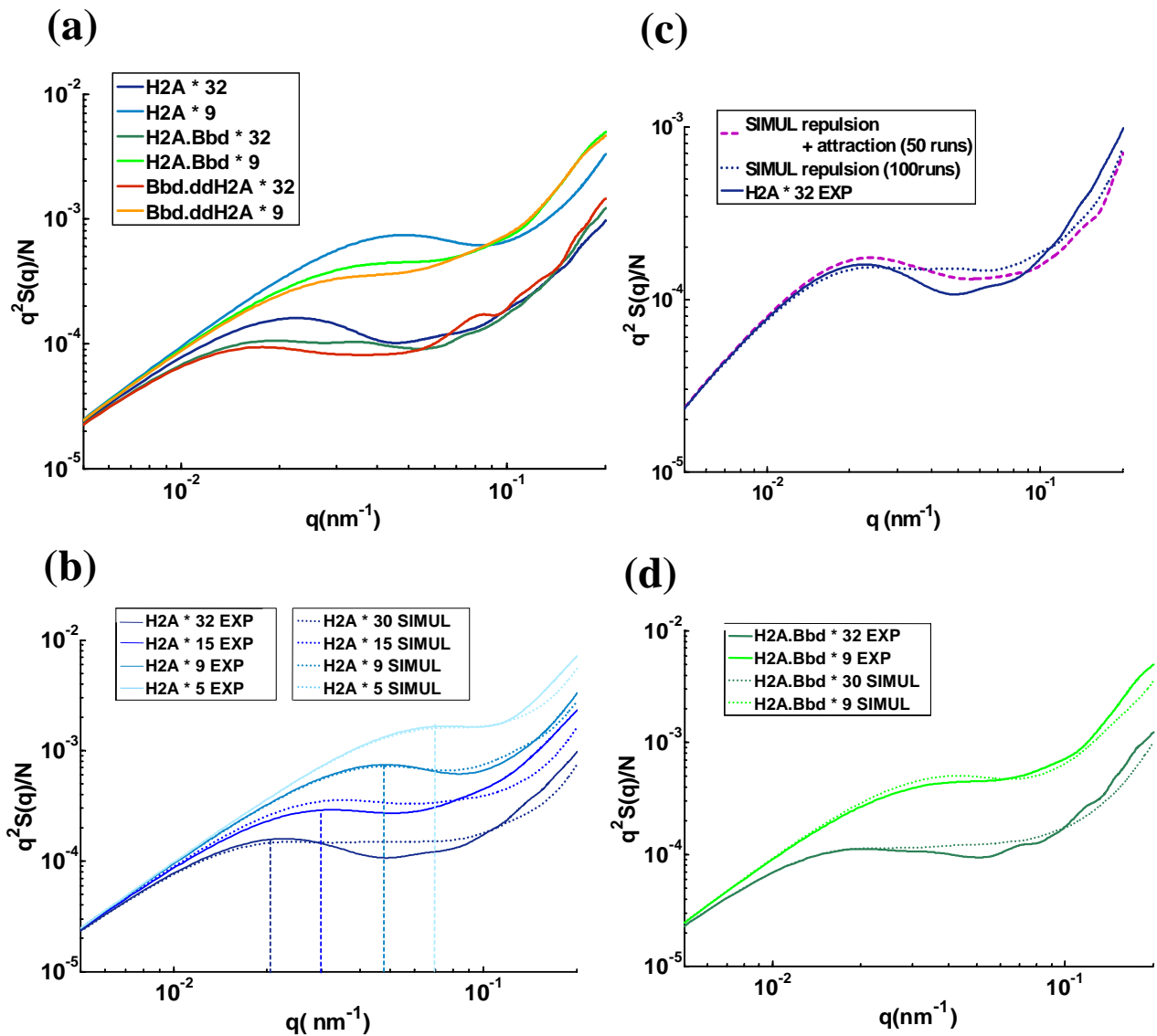